\documentclass[twocolumn]{apex}
\usepackage{txfonts}

\usepackage{graphicx} \usepackage{bm} \usepackage{amsmath}
\usepackage{amssymb} \newcommand{\etal}{{ \it et al. }}

\title{Numerical Study on Spin Torque Switching in Thermally Activated Region}

\author{
  Tomohiro Taniguchi, Mitsutoshi Shibata$^{1}$, Michael Marthaler$^{2}$, Yasuhiro Utsumi$^{1}$, and Hiroshi Imamura} 
\inst{
  Spintronics Research Center, AIST, 1-1-1 Umezono, Tsukuba, Ibaraki 305-8568, Japan, \\
  $^{1}$Faculty of Engineering, Mie University, Tsu, Mie 514-8507, Japan, \\
  $^{2}$Institute for Theoretical Physics, Karlsruhe Institute of Technology, 76128 Karlsruhe, Germany.
}

\abst{
      We studied the spin torque switching of a single free layer in the thermally activated region 
      by numerically solving the Landau-Lifshitz-Gilbert equation. 
      We found that the temperature dependence of the switching time of the in-plane magnetized system is nonlinear, 
      which means $b \neq 1$. 
      Here, $b$ is the exponent of the current term in the switching rate formula
      and has been widely assumed to be unity. 
      This result enables us to evaluate the thermal stability of spintronics devices. 
  }

\begin{document}
\maketitle


Thermal stability is an important property of ferromagnetic materials 
for spintronics device applications such as 
spin random access memory (Spin RAM) and microwave oscillator. 
For example, the greater the thermal stability, 
the longer the Spin RAM retention time \cite{hayakawa08,yakata09}. 
Experimentally, the thermal stability 
is evaluated by measuring the magnetization switching of a free layer 
in the thermally activated region, 
and analyzing the time evolution of the switching probability with the formula \cite{brown63}, 
\begin{equation}
  P(t)
  =
  1 
  -
  \exp
  \left[
    -\int_{0}^{t} 
    \nu(t^{\prime})
    {\rm d}t^{\prime}
  \right],
  \label{eq:switching_probability}
\end{equation}
where $\nu(t)$ is the switching rate given by 
\begin{equation}
  \nu(t)
  =
  f
  \exp
  \left[
    -\Delta_{0}
    \left(
      1
      +
      \frac{H_{\rm appl}}{H_{\rm K}}
    \right)^{2}
    \left(
      1
      -
      \frac{I}{I_{\rm c}}
    \right)^{b}
  \right].
  \label{eq:switching_rate}
\end{equation}
Here, $f$ is the attempt frequency. 
The thermal stability, $\Delta_{0}$, of a uniaxially anisotropic ferromagnet is defined as 
$\Delta_{0}=MH_{\rm K}V/(2k_{\rm B}T)$, 
where $M$, $H_{\rm K}$, $V$, and $T$ are 
the magnetization, 
uniaxial anisotropy field, 
free layer volume, 
and temperature, respectively. 
$I_{\rm c}$ is the critical current of the spin torque switching at zero temperature. 
The exponent of the current term is denoted as $b$. 
Equations (\ref{eq:switching_probability}) and (\ref{eq:switching_rate}) were analytically 
derived for the uniaxially anisotropic system 
by solving the Fokker-Planck equation, 
and $b=2$ as shown in refs. \cite{suzuki09,butler10,taniguchi11}. 
On the other hand, 
for an in-plane magnetized system, 
which has easy and hard axes along and normal to the film plane, respectively, 
and does not have axial symmetry, 
it is difficult to derive the analytical formula of the switching rate 
from the Fokker-Planck equation. 
However, since eq. (\ref{eq:switching_rate}) is the general form of the switching rate 
following the Arrhenius law \cite{hanggi90}, 
eq. (\ref{eq:switching_rate}) has been widely used 
to determine the thermal stability of the in-plane magnetized system \cite{hayakawa08,yakata09}. 

\begin{figure}
\centerline{\includegraphics[width=1.0\columnwidth]{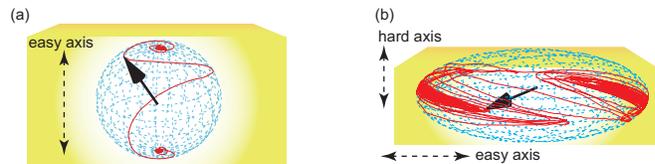}}
\caption{
         (a) Schematic views of the magnetization switching in 
         the uniaxially anisotropic system. 
         The black arrow and red line represent the magnetization and its orbit, respectively. 
         (b) Magnetization switching in the in-plane magnetized system. 
         \vspace{-3ex}}
\label{fig:fig1}
\end{figure}

Let us discuss the differences between 
the uniaxially anisotropic and in-plane magnetized systems 
from the standpoint of the Fokker-Planck approach. 
In the uniaxially anisotropic system,  
the magnetization dynamics shown in Fig. \ref{fig:fig1}(a) is 
described by one variable 
(the angle from the easy axis). 
Moreover, the effect of the spin torque can be included in the effective potential \cite{suzuki09,butler10,taniguchi11}, 
and can be regarded as an additional term to the applied field. 
Then, as the field switching problem \cite{brown63}, 
the switching rate formula can be obtained analytically.
On the other hand, 
in the in-plane magnetized system, 
the magnetization dynamics shown in Fig. \ref{fig:fig1}(b) is described by two angles 
(the angles from the easy and hard axes). 
Also, no effective potential can be introduced 
to describe the spin torque effect. 
These make it difficult to calculate the switching rate analytically. 


In 2004, based on the Landau-Lifshitz-Gilbert (LLG) equation, 
Koch \etal \cite{koch04} showed that $b=1$ 
for the in-plane magnetized system. 
After that, eq. (\ref{eq:switching_probability}) with $b=1$ has been widely used 
to analyze experiments \cite{hayakawa08,yakata09}. 
Recently, however, 
we \cite{taniguchi12c} pointed out that the result in ref. \cite{koch04} 
should be regarded as an approximate one of the uniaxially anisotropic system 
due to the assumption used in their calculation, 
and is not applicable to the in-plane magnetized system. 
Then, the natural question becomes: What is the value of $b$ for the in-plane magnetized system? 
It should be noted that 
the value of $b$ significantly affects 
the evaluation of the quality of the spintronics devices \cite{taniguchi12c}. 
For example, the retention time of the Spin RAM estimated by using the theory with $b=2$ is 
several orders of magnitude longer than that estimated by using $b=1$. 
Thus, the determination of the value of $b$ is important for spintronics. 
In particular, since the in-plane magnetized system is conventionally used 
as the fundamental structure of the Spin RAM, 
the determination of $b$ in this system is an attractive problem 
for practical application.


In this letter, 
we studied the spin torque switching of the single free layer 
in the thermally activated region  
by numerically solving the LLG equation. 
According to our recent studies \cite{taniguchi12a,taniguchi12b}, 
investigation of the temperature dependence of the switching time 
enables us to estimate the value of $b$. 
By comparing these previous works with the current results, 
we found that $b \sim 3$ for in-plane magnetized system. 
This value may not be a universal one. 
However, the important point is that 
our results indicate $b \neq 1$. 



Before proceeding to the in-plane magnetized system, 
we first study the switching of the uniaxially anisotropic system 
to show the consistency between the LLG and Fokker-Planck approaches 
for the spin torque system. 
As mentioned above, 
for the Fokker-Planck approach to the uniaxially anisotropic system, 
the spin torque can be regarded as an additional term to the applied field. 
Thus, the switching probability for $(|h|,|j|)=(\tilde{h},0)$ should be 
identical to that for $(|h|,|j|)=(0,\tilde{h})$, 
where $0 \le \tilde{h} < 1$, 
$h=H_{\rm appl}/H_{\rm K}$, and $j=-I/I_{\rm c}$: see eq. (\ref{eq:switching_rate}).
We used this fact to check the consistency. 


The details of the calculation are as follows. 
We assumed that the magnetization dynamics is described by 
the LLG equation \cite{brown63,slonczewski96,berger96}:
\begin{equation}
  \frac{{\rm d}\mathbf{m}}{{\rm d}t}
  =
  -\gamma
  \mathbf{m}
  \times
  \mathbf{H}
  +
  \gamma
  H_{\rm s}
  \mathbf{m}
  \times
  \left(
    \mathbf{p}
    \times
    \mathbf{m}
  \right)
  -
  \gamma
  \mathbf{m}
  \times
  \mathbf{h}
  +
  \alpha
  \mathbf{m}
  \times
  \frac{{\rm d}\mathbf{m}}{{\rm d}t},
  \label{eq:LLG}
\end{equation}
where $\mathbf{m}$, $\mathbf{H}$, $\gamma$, and $\alpha$ are 
the unit vector pointing in the direction of the magnetization, 
the magnetic field, 
the gyromagnetic ratio, 
and the Gilbert damping constant, respectively. 
$H_{\rm s}=\hbar \eta I/(2eMV)$ and $\mathbf{p}$ are 
the strength of the spin torque 
and the unit vector pointing in the direction of the magnetization of the pinned layer, respectively. 
The positive current with the spin polarization $\eta$ is defined as 
the electron flow from the pinned to the free layer. 
The random field satisfies 
$\langle h_{i}(t)h_{j}(t^{\prime}) \rangle = [2 \alpha k_{\rm B}T/(\gamma MV)]\delta_{ij}\delta(t-t^{\prime})$, 
where $i,j=x,y,z$ \cite{brown63}. 
In eq. (\ref{eq:LLG}), we use the macrospin model 
because the current-resistance curve in ref. \cite{hayakawa08} shows 
rapid changes of the resistance 
between the parallel ($\mathbf{m}=\mathbf{p}$) and antiparallel ($\mathbf{m}=-\mathbf{p}$) alignments, 
which means a uniform rotation of the magnetization. 
We assume that both the easy axis of the free layer and $\mathbf{p}$ are parallel to the $z$-axis 
($\mathbf{p}=(0,0,1)$). 
The initial state is set to be $\mathbf{m}(0)=(0,0,1)$. 
The 4${}^{\rm th}$-order Runge-Kutta method was employed to solve eq. (\ref{eq:LLG}). 
The magnetization dynamics is averaged over $10^{5}$ samples. 
The switching probability 
is obtained by counting the sample numbers in which $m_{z} \le -0.9$. 
Once the magnetization reaches $m_{z} \le -0.9$, 
we regard the sample as the switched system, 
even if the magnetization returns to the region $m_{z}>-0.9$ 
due to the thermal fluctuations. 



\begin{figure}
\centerline{\includegraphics[width=1.0\columnwidth]{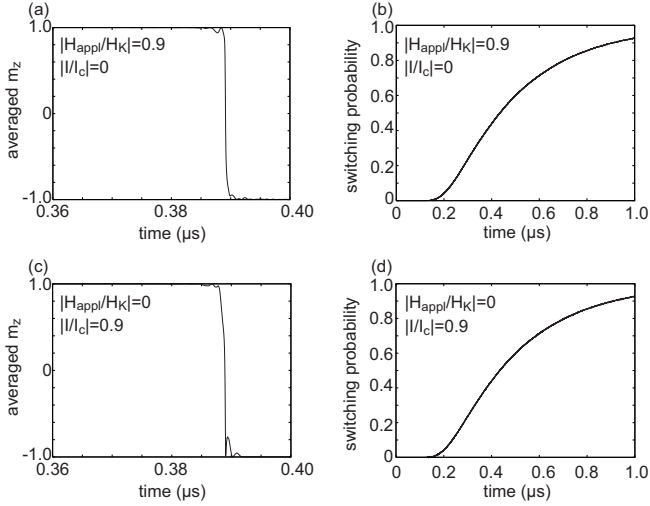}}
\caption{
         Time evolutions of (a) the averaged $m_{z}$ and (b) the switching probability 
         under the effect of the applied field only. 
         The system is uniaxially anisotropic. 
         The time evolutions of (c) the averaged $m_{z}$ and (d) the switching probability 
         under the effect of the spin torque only are also shown. 
         \vspace{-3ex}}
\label{fig:fig2}
\end{figure}



For the uniaxially anisotropic system, 
the magnetic field is given by $\mathbf{H}=(0,0,H_{\rm appl}+H_{\rm K}m_{z})$. 
The critical current 
from the parallel to the anti-parallel alignment 
is $I_{\rm c}=-[2\alpha eMV/(\hbar\eta)](H_{\rm appl}+H_{\rm K})$.
According to the definition, 
the negative $h$ and $j$ favor the antiparallel alignment. 
Figures \ref{fig:fig2}(a) and \ref{fig:fig2}(b) show 
the time evolutions of the averaged $m_{z}$ and switching probability 
under the effect of the applied field only ($j=0$), respectively.  
The values of the parameters are taken to be 
$M=1000$ emu/c.c., $H_{\rm K}=200$ Oe, $\gamma=17.64$ MHz/Oe, $\alpha=0.01$, and $T=300$ K. 
The thickness and cross-sectional area of the free layer are 
taken to be $2.5$ nm and $\pi \times 80 \times 35$ nm$^{2}$, respectively \cite{yakata09}. 
The magnitude of the applied field, $H_{\rm appl}=-180$ Oe, 
is 90\% of the anisotropy field ($h=-0.9$). 
On the other hand, in Figs. \ref{fig:fig2}(c) and \ref{fig:fig2}(d), 
the time evolutions of the averaged $m_{z}$ and switching probability 
under the effect of the spin torque only ($h=0$) are shown, 
where $j=-0.9$. 
The switching probability 
for $(|h|,|j|)=(0.9,0)$ is 
almost identical to 
that for $(|h|,|j|)=(0,0.9)$. 
This result indicates the consistency between the LLG and Fokker-Planck approaches 
for the spin torque system, 
and can be regarded as evidence of $b=2$ 
from the LLG approach.


Figures \ref{fig:fig2}(b) and \ref{fig:fig2}(d) show the waiting time during which the switching probability remains zero, 
while eq. (\ref{eq:switching_probability}) predicts a slight increase of the switching probability from $t=0$. 
The waiting time was found in both experiments \cite{tomita08} and 
numerical calculation of the Fokker-Planck equation \cite{butler10}. 
Although the waiting time is important for device application, 
further study is beyond the scope of this letter. 


Next, we move to the in-plane magnetized system. 
The magnetic field is given by 
$\mathbf{H}=(0,-4\pi Mm_{y},H_{\rm appl}+H_{\rm K}m_{z})$, 
where the $y$-component includes the demagnetization field. 
In this system, 
the value of $b$ can be estimated as follows. 
By assuming the constant attempt frequency $f_{0}$ 
and the sweep current $I(t)=\varkappa t$ with the sweep rate $\varkappa$ 
as is done in the experiments \cite{albert02}, 
eqs. (\ref{eq:switching_probability}) and (\ref{eq:switching_rate}) with $H_{\rm appl}=0$ reduce to 
\begin{equation}
\begin{split}
  P(t)
  =
  1
  -
  \exp
  &
  \left\{
    -\frac{f_{0}I_{\rm c}}{b \varkappa \Delta_{0}^{1/b}}
    \left[
      \gamma
      \left(
        \frac{1}{b},
        \Delta_{0}
      \right)
      -
      \gamma
      \left(
        \frac{1}{b},
        \Delta_{0}
        \left(
          1 - \frac{\varkappa t}{I_{\rm c}}
        \right)^{b}
      \right)
    \right]
  \right\},
\end{split}
\end{equation}
where $\gamma(\beta,z)$ is the lower incomplete $\Gamma$ function. 
As shown in ref. \cite{taniguchi12c},
the probability density, ${\rm d} P(t)/{\rm d} t$, has its maximum at a certain time $\tilde{t}$. 
We call $\tilde{t}$ the switching time, 
which is given by 
\begin{equation}
  \tilde{t}
  =
  \frac{I_{\rm c}}{\varkappa}
  \left[
    1
    -
    \frac{1}{\Delta_{0}}
    \log 
    \left(
      \frac{f_{0}I_{\rm c}}{\varkappa \Delta_{0}}
    \right)
  \right],
  \label{eq:switching_time_b_1}
\end{equation}
for $b=1$, 
and 
\begin{equation}
  \tilde{t}
  =
  \frac{I_{\rm c}}{\varkappa}
  \left(
    1
    -
    \left\{
      \frac{b-1}{b\Delta_{0}}
      {\rm plog}
      \left[
        \frac{b}{b-1}
        \left(
          \frac{f_{0}I_{\rm c}}{b \varkappa \Delta_{0}^{1/b}}
        \right)^{b/(b-1)}
      \right]
    \right\}^{1/b}
  \right),
  \label{eq:switching_time_b_arb}
\end{equation}
for $b>1$. 
Here, ${\rm plog}(z)$ is the product logarithm. 
By assuming that $\Delta_{0} \propto 1/T$, 
in the low-temperature region, 
the temperature dependence of $\tilde{t}$ is approximately linear if $b=1$, 
whereas it is nonlinear if $b>1$ \cite{taniguchi12a,taniguchi12b}. 
Thus, we investigated $\tilde{t}$ 
by solving eq. (\ref{eq:LLG}), 
and compared it with eq. (\ref{eq:switching_time_b_1}) or (\ref{eq:switching_time_b_arb}). 


In the calculation, 
the current sweep rate is taken to be 30 A/s.
The switching time is obtained 
by fitting the probability density with a Gaussian curve. 
It should be noted that 
the magnitude of the critical current, 
$I_{\rm c}$, in eq. (\ref{eq:switching_time_b_1}) or (\ref{eq:switching_time_b_arb}) is 
usually larger than that estimated on the basis of 
the instability of the initial state \cite{sun00},
$-[2\alpha e MV/(\hbar\eta)](H_{\rm K}+2\pi M)=-0.54$ mA 
($\eta$ is taken to be $0.8$),
because the instability of the initial state 
does not guarantee switching. 
Thus, the switching time shown below is longer than 
$[2\alpha e MV/(\varkappa\hbar\eta)](H_{\rm K}+2\pi M)=18$ $\mu$s, 
and eq. (\ref{eq:switching_time_b_1}) or (\ref{eq:switching_time_b_arb}) has 
four fitting parameters, $I_{\rm c}$, $b$, $\Delta_{0}T,$ and $f_{0}$.



\begin{figure}
\centerline{
\includegraphics[width=0.7\columnwidth]{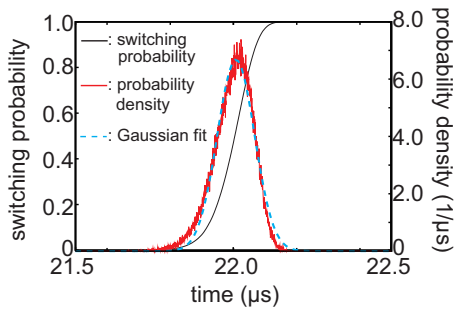}}
\caption{
         Time evolutions of the switching probability of the in-plane magnetized system (black) 
         and the probability density (red) at $T=1$ K.
         The dotted line (blue) is the Gaussian fit, where 
         the central time is defined as the switching time.
         \vspace{-3ex}}
\label{fig:fig3}
\end{figure}



\begin{figure}
\centerline{
\includegraphics[width=1.0\columnwidth]{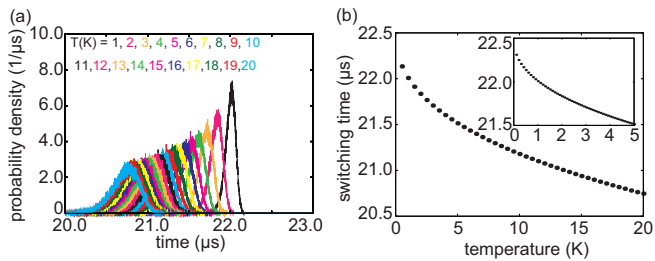}}
\caption{
         (a) Probability densities for the various temperatures.
         (b) Temperature dependence of the switching time of the in-plane magnetized system. 
         The inset shows the switching time in the very low temperature region. 
         \vspace{-3ex}}
\label{fig:fig4}
\end{figure}

Figure \ref{fig:fig3} shows the time evolutions of 
the switching probability and probability density 
at $T=1$ K. 
The probability density can be well reproduced by the Gaussian curve 
as indicated by the dotted line. 
The probability densities for various temperatures ($\le 20$ K) 
and their switching time are shown in Figs. \ref{fig:fig4}(a) and \ref{fig:fig4}(b), respectively. 
The switching time shows 
nonlinear temperature dependence, 
which means $b \neq 1$. 
While the estimated values of $\Delta_{0}T$ and $f_{0}$ depend heavily on 
the initial values of the fitting, 
the estimated value of $b$ depends weakly on the initial value, 
and is $b \sim 3$. 



In an infinite demagnetization field limit, 
the switching is completely limited in the in-plane, 
i.e., one-dimensional motion as the uniaxially anisotropic system, 
and $b=2$ \cite{taniguchi11}. 
In a realistic system, 
while there are infinite numbers of possible switching paths, 
the averaged magnetization dynamics is determined mainly by 
the path along the in-plane:
see Fig. \ref{fig:fig1}(b).
However, 
because the demagnetization field is finite, 
the magnetization can move to the hard axis direction. 
Due to the contribution of such paths, 
$b$ should deviate from 2. 
In other words, our result may indicate that 
$b$ of the in-plane magnetized system depends on 
the material parameters, such as the demagnetization coefficient. 
Although the investigation on such dependence is under consideration, 
the important point is that our results indicate $b \neq 1$ 
for the in-plane magnetized system. 
We also briefly consider 
why $b$ estimated above is larger than 2. 
Compared with the infinite demagnetization field limit, 
the realistic system has a large number of switching paths, 
and thus, a high switching probability. 
Since the large $b$ corresponds to a large number of switching events, 
$b$ may be larger than 2. 
However, more accurate discussion requires the analytical study based on 
the Fokker-Planck equation, and is beyond the scope of this letter.  


In summary, we studied the spin torque dependence of the magnetization switching probability 
by numerically solving the Landau-Lifshitz-Gilbert equation. 
We found that the temperature dependence of the switching time is nonlinear, 
i.e., $b \neq 1$, 
where $b$ is the exponent of the current term in the switching rate formula. 
The results will be important for evaluating the thermal stability of Spin RAM.


The authors would like to acknowledge 
S. Yuasa, 
H. Kubota, 
Y. Suzuki, 
H. Sukegawa, 
and S. Mitani 
for the valuable discussions they had with us. 




\begin{thebibliography}{25}
%
\bibitem{hayakawa08}
J. Hayakawa, S. Ikeda, K. Miura, M. Yamanouchi, Y. M. Lee, R. Sasaki,
M. Ichimura, K. Ito, T. Kawahara, R. Takemura, T. Meguro,
F. Matsukura, H. Takahashi, H. Matsuoka, and H. Ohno: 
IEEE. Trans. Magn. {\bf 44} (2008) 1962.

\bibitem{yakata09}
S. Yakata, H. Kubota, T. Sugano, T. Seki, K. Yakushiji, A. Fukushima, S. Yuasa, and K. Ando: 
Appl. Phys. Lett. {\bf 95} (2009) 242504.

\bibitem{brown63}
W. F. Brown Jr.: Phys. Rev. {\bf 130} (1963) 1677.

\bibitem{suzuki09}
Y. Suzuki, A. A. Tulapurkar, and C. Chappert: 
\textit{Nanomagnetism and Spintronics} (Elsevier, Amsterdam, 2009) 1st ed., Chap. 3.

\bibitem{butler10}
W. H. Butler: presented at 55$^{\rm th}$ Annual Conference on Magnetism and Magnetic Materials, 2010.
See also arXiv:1202.2621. 

\bibitem{taniguchi11}
T. Taniguchi and H. Imamura: Phys. Rev. B {\bf 83} (2011) 054432.

\bibitem{hanggi90}
P. H\"anggi, P. Talkner, and M. Borkovec: Rev. Mod. Phys. {\bf 62} (1990) 251.

\bibitem{koch04}
R. H. Koch, J. A. Katine, and J. Z. Sun: Phys. Rev. Lett. {\bf 92} (2004) 088302. 

\bibitem{taniguchi12c}
T. Taniguchi and H. Imamura: 
to be published in IEEE Trans. Magn. 
See arXiv:1204.1190.

\bibitem{taniguchi12a}
T. Taniguchi and H. Imamura: 
to be published in J. Nanosci. Nanotechnol.
See arXiv:1204.2004.

\bibitem{taniguchi12b}
T. Taniguchi and H. Imamura: 
J. Appl. Phys. {\bf 111} (2012) 07C901.

\bibitem{slonczewski96}
J. C. Slonczewski: J. Magn. Magn. Mater. {\bf 159} (1996) L1.

\bibitem{berger96}
L. Berger: Phys. Rev. B {\bf 54} (1996) 9353.

\bibitem{tomita08}
H. Tomita, K. Konishi, T. Nozaki, H. Kubota, A. Fukushima, K. Yakushiji, S. Yuasa, Y. Nakatani, T. Shinjo, M. Shiraishi, and Y. Suzuki: 
Appl. Phys. Express {\bf 1} (2008) 061303. 

\bibitem{albert02}
F. J. Albert, N. C. Emley, E. B. Myers, D. C. Ralph, and R. A. Buhrman: 
Phys. Rev. Lett. {\bf 89} (2002) 226802.

\bibitem{sun00}
J. Z. Sun: 
Phys. Rev. B {\bf 62} (2000) 570.

\end{thebibliography}
\end{document}